\def\flash{1}
\def\aa{2}
\def\jina{3}
\def\hebrew{4}
\def\efi{5}
\def\anl{6}

\def\rb{r_{\rm bubble}}
\def\ro{r_{\rm offset}}
\def\lc{\lambda_c}
\def\vjet{v_{\rm jet}}
\def\enuc{E_{\rm nuc}}

\def\tmax{T_{\rm max}}

\documentclass[preprint,12pt]{aastex}
\newlength{\GBCdigit}
\newlength{\GBCminus}
\shorttitle{Three-Dimensional Simulations of Type Ia Supernovae}
\shortauthors{Jordan et al.}

\begin{document}

\title{Three-Dimensional Simulations of the Deflagration Phase
of the Gravitationally Confined Detonation Model of Type Ia Supernovae}

\author{
G.~C.~Jordan IV,\altaffilmark{\flash,\aa}
R.~T.~Fisher,\altaffilmark{\flash,\aa}
D.~M.~Townsley,\altaffilmark{\jina,\aa}
A.~C.~Calder,\altaffilmark{\flash,\aa}
C.~Graziani,\altaffilmark{\flash,\aa}
S.~Asida,\altaffilmark{\hebrew}
D.~Q.~Lamb,\altaffilmark{\flash,\aa,\efi}
~and~J.~W.~Truran,\altaffilmark{\flash,\aa,\efi,\anl}
}
\altaffiltext{\flash}{Center for Atrophysical Thermonuclear Flashes, The
University of Chicago, Chicago, IL 60637.}

\altaffiltext{\aa}{Department of Astronomy and Astrophysics, The
University of Chicago, Chicago, IL 60637.}

\altaffiltext{\jina}{Joint Institute for Nuclear Astrophysics, The
University of Chicago, Chicago, IL 60637.}

\altaffiltext{\hebrew}{Racah Institute of Physics, Hebrew University,
Jerusalem 91904, Israel.}

\altaffiltext{\efi}{Enrico Fermi Institute, The University of Chicago,
Chicago, IL 60637.}

\altaffiltext{\anl}{Argonne National Laboratory, Argonne, IL 60439.}

\begin{abstract}
We report the results of a series of three-dimensional (3-D)
simulations of the deflagration phase of the gravitationally confined
detonation mechanism for Type Ia supernovae.  In this mechanism,
ignition occurs at one or several off-center points, resulting in a
burning bubble of hot ash that rises rapidly, breaks through the
surface of the star, and collides at a point opposite breakout on the
stellar surface.  We find that detonation conditions are robustly
reached in our 3-D simulations for a range of initial conditions and
resolutions.  Detonation conditions are achieved as the result of an
inwardly-directed jet that is produced by the compression of unburnt
surface material when the surface flow collides with itself. A
high-velocity outwardly-directed jet is also produced.  The initial
conditions explored in this paper lead to conditions at detonation
that can be expected to produce large amounts of $^{56}$Ni and small
amounts of intermediate mass elements.  These particular simulations
are therefore relevant only to high luminosity Type Ia supernovae. 
Recent observations of Type Ia supernovae imply a compositional
structure that is qualitatively consistent with that expected from
these simulations.
\end{abstract}


\section{Introduction}

Type Ia supernovae have received increased interest because of their
importance as ``standard candles'' for cosmology.  Observations using
Type Ia supernovae as standard candles have revealed that the expansion
rate of the universe is accelerating and have led to the discovery of
``dark energy'' \citep{riess1998,perlmutter1998}.  But the way in which
Type Ia supernovae explode is not completely understood.  The current
leading paradigms for the explosion mechanism are (1) pure deflagration
\citep{reinecke2002b,gamezo2003,roepke2005}, (2) deflagration to
detonation transition (DDT) \citep{khokhlov1991,gamezo2004,gamezo2005},
(3) pulsational detonation (PD) \citep{khokhlov1991,bravo2006}, and (4)
gravitationally confined detonation (GCD)
\citep{plewa2004,livne2005,plewa2007,townsley2007}.  There is
increasing evidence that a detonation is needed
\citep{hoeflich2002,badenes2006,wang2006,wang2007,gerardy2007}, as is
posited in the last three models.

A fundamental question has been how the transition to a detonation
occurs in a white dwarf star [see, e.g., \citet{niemeyer1999}].  While
the DDT, PD, and GCD paradigms incorporate a detonation, all existing
DDT simulations invoke the transition to a detonation in an ad-hoc
fashion and the PD mechanism remains largely unexplored by detailed
simulations.  In contrast, extensive two-dimensional (2-D) cylindrical
simulations have shown that detonation conditions are robustly reached
in the GCD model for a variety of initial conditions
\citep{plewa2004,plewa2007,roepke2007,townsley2007}.  Thus, to date,
the GCD mechanism is the only proposed mechanism for which a
detonation has been demonstrated to arise naturally.

However, the achievement of detonation conditions has not been
demonstrated in 3-D [see, e.g., \citet{roepke2007}].  This is a concern
since the behavior of turbulence is different in 3-D than in 2-D, and
the cylindrical symmetry of the 2-D simulations might enhance the
focusing of the surface flow that triggers the detonation in the GCD
model.  Hence, a major question that we address in this paper is
whether it is possible to achieve detonation conditions in a fully 3-D
simulation of the GCD model.  In addition, the large amount of nuclear
burning that occurs in the \citet{roepke2007} simulations appears to 
play a role in the failure of these simulations to achieve detonation
conditions in 3-D.  In this paper, we report the results of seven 3-D
simulations of the GCD model for several different sets of initial
conditions and two resolutions.  We find that the conditions for
detonation are robustly achieved for these initial conditions.  Thus
the simulations reported in this paper address the first point above
and provide a counter example to the second.

The organization of the paper is as follows.  We describe the
simulation setup in \S 2 and the results of the simulations in \S 3. We
discuss the properties of Type Ia supernovae expected in the GCD model
and compare the results of our simulations with earlier work in \S 4. 
Finally, we state our conclusions in \S 5.

\section{Simulation Setup} \label{simulation_setup}

We perform our 3-D simulations of the deflagration phase of the GCD
model using FLASH 3.0, an adaptive-mesh hydrodynamics code
\citep{fryxell2000,calder2002}.  The general simulation setup is
identical to that in the 2-D simulations of the GCD model described in
\citet{townsley2007}.  In particular, the initial model is a 1.38
$M_\sun$ WD with a uniform composition of equal parts by mass of
$^{12}$C and $^{16}$O.  It has a central density of $2.2 \times 10^9$ g
cm$^{-3}$, a uniform temperature of $4 \times 10^7$ K, and a radius of
approximately 2,000 km.  

The physical thickness of the carbon burning stage of the nuclear flame
front in the deflagration phase of Type Ia supernovae is $10^{-4}$ to
$10^3$ cm for the densities of interest, and is therefore unresolvable
in any whole-star simulation.  Consequently, a method must be used to
determine the location and speed of the flame front.  Two fairly
different methods of flame-front tracking have been used in recent
studies of WD deflagration.  One is the level set technique, in which
the location of the flame front is calculated based upon the value of a
smooth field defined on the grid and propagated with an advection
equation acting in addition to the hydrodynamics
\citep{reinecke1999,roepke2003,roepke2005}.  This method has been used
to study the effect of turbulence on the nuclear burning rate
\citep{schmidt2006a} and in many simulations of WD deflagration [see,
e.g., \citet{reinecke1999,roepke2005,schmidt2006b}].  The other method
artificially broadens the flame front using a reaction progress
variable and propagates it using an advection-diffusion-reaction (ADR)
equation \citep{khokhlov1995,vladimirova2006}.  This ADR flame model
has been used to study the effect of the Rayleigh-Taylor (R-T)
instability on a propagating flame front \citep{khokhlov1995,zhang2007}
and in many previous previous simulations of WD deflagration
\citep{gamezo2003, calder2004,plewa2004,plewa2007,townsley2007}.

In this paper, we follow the nuclear flame using a new version
\citep{asida2008} of ADR flame model.  The new prescription uses the
Kolmogorov-Petrovski-Piskunov (KPP) form of the reaction term in which
this term is slightly truncated, as opposed to the top-hat form used
previously by ourselves and others [e.g., \citet{khokhlov1995}].  This
new version is numerically quieter, more stable, and exhibits far
smaller curvature effects \citep{asida2008}.  The thickness of the
flame is $\approx 4$ grid points; more details and the explicit values
of the parameters in the ADR flame model that we use are given in
\citet{townsley2007}.  We also use a new, acoustically-quiet version
\citep{townsley2007} of the nuclear energy release method described in
\cite{calder2007} that accounts more accurately for the nuclear energy
released in the flame and in the evolution of nuclear statistical
equilibrium (NSE) as conditions change within the bubble of hot ash.

We do not include nuclear burning outside the flame in the simulations
reported in this paper.  This approach is the same as that adopted by
\citet{roepke2007}.  Were we to have included such burning, both the
2-D simulations reported in \citet{townsley2007} and the 3-D
simulations reported in this paper would detonate at the first instant
at which detonation conditions are reached, as we will report in later
papers \citep{meakin2008a,meakin2008b}.  Thus including nuclear burning
outside the flame would not have allowed us to demonstrate that the
simulations reported in this paper  robustly reach conservative
conditions for detonation; i.e., that they exceed the temperature and
the density needed to detonate for a significant period of time.

We treat the effect of R-T--driven turbulence in the same manner as in
\citet{townsley2007}; namely, we impose a minimum flame speed $s_{\rm
min} = \alpha \sqrt{A g m_f \Delta x}$, where $\Delta x$ is the grid
size of the simulation, $\alpha = 0.5$ is a geometrical factor, and
$m_f = 0.06$ is a constant that we have calibrated
\citep{townsley2007,townsley2008}.  This is a conservative approach
that allows the simulation to treat the effects of R-T driven
turbulence on resolved scales while ensuring that turbulence on the
scale of the (artificially broadened) flame thickness does not disrupt
the flame.  We disable this prescription in the truncated cone
encompassing the region where the surface flow collides with itself to
ensure that no unrealistic heating occurs.

If the nuclear burning rate, as well as the overall dynamics of
R-T--driven turbulent nuclear burning, are determined by the behavior
at large scales, turbulence at small scales does not increase the
burning rate and the moderate resolution possible in current
simulations is adequate.  That this may be the case is suggested by the
results of \cite{zhang2007}, who find that the time-averaged rate of
buoyancy-driven nuclear burning did not vary when the resolution was
varied by a factor of four.  Because we rely on the resolution of our
simulations to describe turbulent nuclear burning, and therefore the
rate of nuclear burning, we have paid close attention to how the rate
varies with resolution.  We have found no evidence for more than a
modest variation of the nuclear burning rate due to unresolved 
behavior -- as we discuss below.  However, the appropriate way to treat
turbulent nuclear burning is an open question [cf.
\citet{reinecke2002a,zingale2005,schmidt2006a,zhang2007}], and further
studies are needed in order to answer definitively this question.  

The core of the star is thought to be convective and therefore to have
velocities on the largest scales of $\sim 100$ km s$^{-1}$
\citep{woosley2004,wunsch2004}, which are comparable to the laminar
flame speed in this region \citep{timmes1992}.  The temperature
fluctuations in the convective region may lead to one, to a few, or to
many ignition points.  The convective motions are likely to distort the
ignition region(s), seeding later Rayleigh-Taylor instability modes. 
However, the motions are not strong enough to destroy the ignition
region(s), once born.  In addition, \citet{livne2005} find that the
general outcome of off-center ignitions is not strongly affected by the
presence of a convective velocity field.  Finally, we would like to
understand the behavior of the simpler case of a single, spherical
ignition region and zero velocity in the core of the star before
considering more complicated cases.  For all of these reasons, we adopt
a single ignition point at a range of offset distances from the center
of the star as the initial conditions in this paper.  We will
investigate ignition at multiple points and the effects of convective
motions in the core of the star in future papers.

We model the ignition region as a spherical bubble of hot ash initially
at rest, characterized by an initial radius, $\rb$, and an initial
distance, $\ro$, from its center to the center of the star along the
z-axis.  The edge of the burned region forms a smooth transition from
fuel to ash that is $\sim$ 4 zones in width [see \citet{townsley2007}
for more details].  The density of the hot ash is chosen to maintain
pressure equilibrium with the surrounding material.  The surface of the
bubble of hot ash corresponds to the $\phi_1 = 0.5$ isosurface, where
$\phi_1$ is the flame progress variable for $^{12}$C $+$ $^{12}$C
burning \citep{townsley2007}.  Thus the radius, $\rb$, is
approximately the radius of the $\phi_1 = 0.5$ isosurface.

Initially, the spherical bubble of hot ash rises slowly, due to its
small size and the small value of the acceleration of gravity $g$ near
the center of the star.  The growth of the bubble is self-similar --
i.e., independent of its initial radius -- provided that $\rb \lesssim
\lc \equiv 6 \pi s^2/Ag$  \citep{vladimirova2007,fisher2008}.  Here
$\rb$ is the initial radius of the bubble, $\lc$ is the minimum
wavelength for the unstable Rayleigh-Taylor growth of flame surface
perturbations, $s$ is the laminar flame speed, $A = (\rho_{\rm fuel} -
\rho_{\rm ash})/(\rho_{\rm fuel} + \rho_{\rm ash})$ is the Atwood
number, and $g$ is the acceleration of gravity at $\ro$.  Otherwise,
the initial bubble is immediately unstable to the growth of
Rayleigh-Taylor modes.  This is inconsistent with the assumption of a
small, spherical ignition region.  Thus, $\rb \lesssim \lc$ for all of
the simulations reported in this paper, except for two we conducted to
compare with those of \cite{roepke2007}.

The adaptive mesh refinement criteria we use are chosen to capture the
relevant physical features of the burning and the flow at reasonable
computation expense.  The criteria for refinement are the same as those
described in \citet{townsley2007}, except that we maximally refine a
sphere of radius $r_{\rm refine} = 1000$ km at the center of the star
and a truncated cone encompassing the region where the flow of hot
bubble material over the surface of the star collides with itself.  The
truncated cone has an opening half angle $\theta_{\rm cone} = 30^o$,
and extends from a radius $r_{\rm cone,min} = 1500$ km to a radius
$r_{\rm cone,max} = 3000$ km.

\section{Results} \label{results}

We performed a suite of seven 3-D simulations of the deflagration phase
of the GCD model for initially stationary, spherical flame bubbles,
varying the initial bubble radius $\rb$, the initial bubble offset
$\ro$, and the finest resolution.  Table 1 lists the initial conditions
and several properties of these simulations.  The simulations are
denoted by initial bubble radius, offset distance, and finest 
resolution.  Thus 18b42o6r denotes a 3-D simulation in which the initial
bubble radius $\rb = 18$ km, the offset distance $\ro = 42$ km, and the
finest resolution is 6 km.

Our 3-D simulations of the GCD model progress similarly to previous 2-D
GCD simulations \citep{plewa2004,plewa2007,townsley2007}, passing
through several distinct stages.  Initially, the spherical bubble of
hot ash that we adopt as our model of the ignition region grows at a
rate dictated by the laminar flame speed.  At $\sim 0.2-0.3$ s of
simulation time, the radius $\rb$ of the bubble exceeds $\lc$, the
minimum wavelength for the unstable Rayleigh-Taylor growth of flame
surface perturbations \citep{khokhlov1995,zhang2007}.  When this
happens, the top surface of the bubble develops a bulge, and the bubble
quickly evolves into a mushroom-like shape
\citep{calder2004,plewa2004,vladimirova2007,plewa2007,townsley2007}. 
Subsequently, the shape of the bubble becomes ever more complex as the
critical wavelength $\lc$ becomes smaller (as the bubble rises and $g$
at the position of the bubble increases) and additional generations of
smaller features appear as a result of the Rayleigh-Taylor
instability.  During this time, the rate at which the bubble rises 
increases, and the bubble breaks through the stellar surface at $\sim
0.8-1.2$ s.  The hot ash in the bubble (which was produced at a range
of densities and so consists of both iron-peak and intermediate-mass
elements) then spreads rapidly over the surface of the star, pushing
unburned material in the outermost layer of the star ahead of it.

The mass of the hot ash from the bubble that sweeps over the surface of
the star ranges from 0.038-0.010 $M_\sun$ for offset distances of the
initial bubble ranging from 20 km to 100 km.  Its composition ranges
from roughly 0.009-0.002 $M_\sun$ of intermediate mass elements and
roughly 0.007-0.03 $M_\sun$ of Fe-peak elements for these offset
distances.  However, we caution that the present simulations do not
seek to accurately treat mixing between the ash flow across the surface
of the star and the underlying outermost layers of the white dwarf,
which will significantly affect the final composition of these layers.

\begin{figure}
\begin{center}
\includegraphics[scale=0.46,clip=]{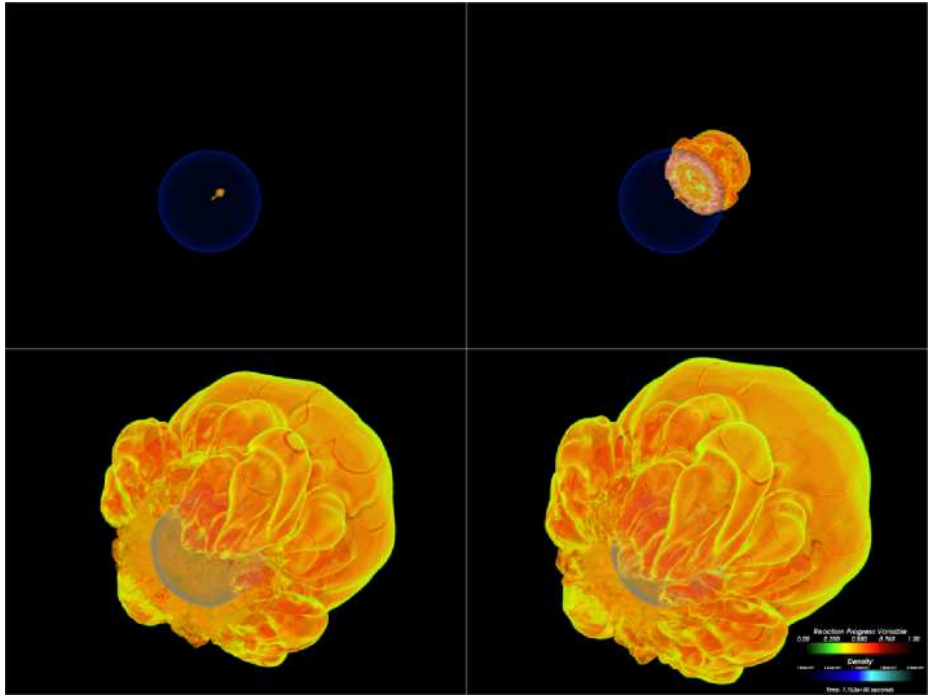}
\end{center}
\caption{Images showing the hot ash and the star at different times for
the 6-km resolution simulation of the GCD model starting from initial
conditions in which a 25-km radius initial bubble offset 100 km from the
center of the star.  The images show volume renderings of the surface of
the star (defined as the region in which $\rho = (1.5-2.0) \times 10^7$
g cm$^{-3}$ and the flame surface (defined as the surface where the
flame progress variable $\phi_1 = 0.5$) at (a) 0.5 s, soon after the
bubble becomes Rayleigh-Taylor unstable and develops into a mushroom
shape, (b) 1.0 s, as the bubble breaks through the surface of the star,
and (c) 1.5 s, when the hot ash is flowing over the surface of the star,
and (d) 1.7 s, shortly before the hot ash from the bubble collides at
the opposite point on the surface of the star
\label{fig:flame_surface}}
\end{figure}

\begin{figure}
\begin{center}
\includegraphics[scale=0.46,clip=]{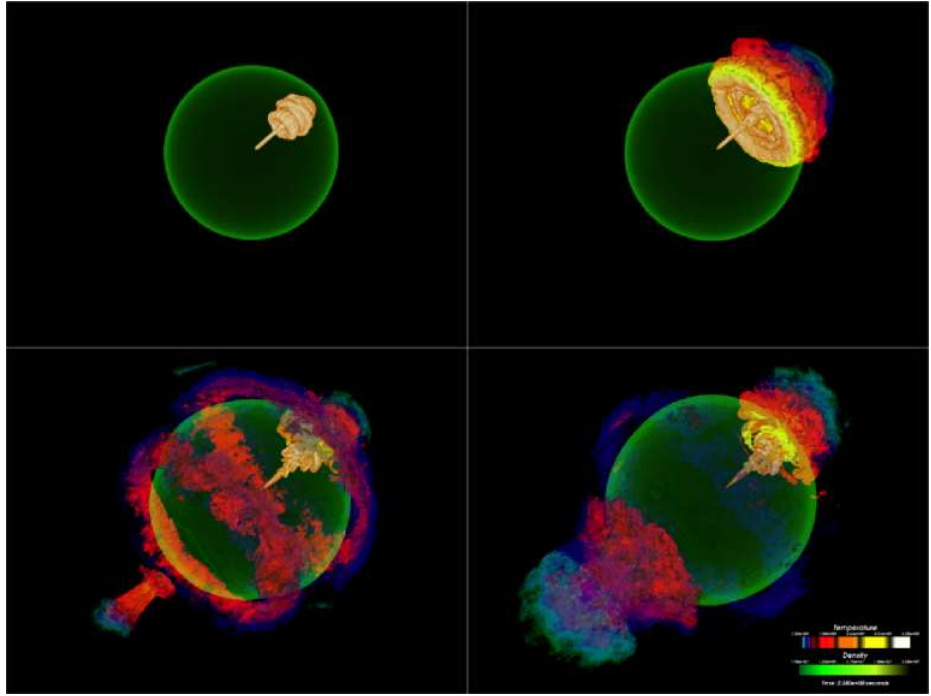}
\end{center}
\caption{Images showing very hot matter and the star at different times
for the same simulation as in Figure \ref{fig:flame_surface}.  The
images are volume renderings of the surface of the star (defined as the
region in which $\rho = (1.5-2.0) \times 10^7$ g cm$^{-3}$ and the
regions where the temperature is very high (i.e., where $T > 1.5 \times
10^9$ K) at (a) 0.8 s, when the bubble has become Rayleigh-Taylor
unstable and developed into a mushroom shape; (b) 1.0 s, as the bubble
breaks through the surface of the star; (c) 1.7 s, shortly before the
hot ash from the bubble collides at the opposite point on the surface of
the star, and (d) 1.84 s, the moment when the inward jet has compressed
and heated stellar material ahead of it to detonation conditions.
\label{fig:hot_regions}}
\end{figure}

\begin{figure}
\begin{center}
\includegraphics[scale=0.5,clip=]{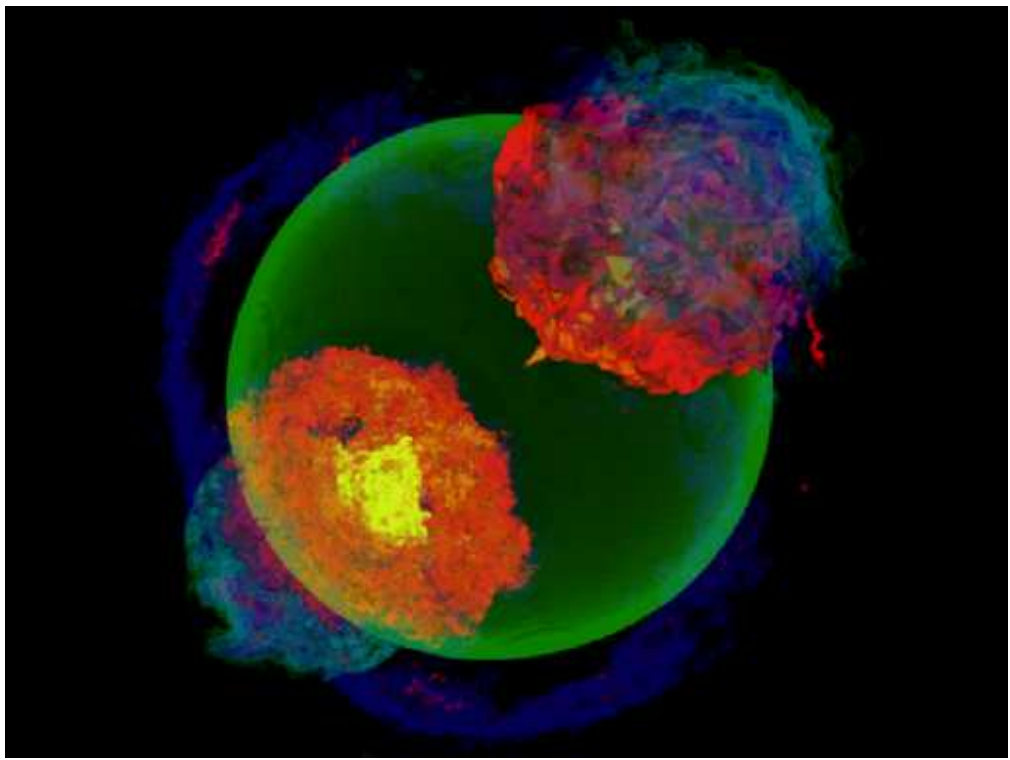}
\end{center} \caption{Same image as that in Figure 2(d), except we have
rotated the star to show the face of the inwardly directed jet where
conditions for detonation are robustly achieved.
\label{fig:hot_regions_rotated}}
\end{figure}

At $\sim 1.8-2.2$ sec, the flow of hot ash collides with itself at the
opposite point on the stellar surface from the place where the bubble
broke out, compresses the unburnt surface layers there, and initiates a
detonation \citep{plewa2004,plewa2007,townsley2007}.  Figures
\ref{fig:flame_surface} and \ref{fig:hot_regions} illustrate these
different stages in the 18b42o6r simulation.  The image in Figure 
\ref{fig:hot_regions_rotated} is identical to that in Figure
\ref{fig:hot_regions}(d), except we have rotated the star to show the
very hot ($T > 3 \times 10^9$ K) region at the head of the inward jet
that has reached densities greater than $2 \times 10^7$ g cm$^{-3}$,
conditions that exceed conservative criteria for initiation of a
detonation \citep{niemeyer1997,roepke2007}.

\begin{figure}
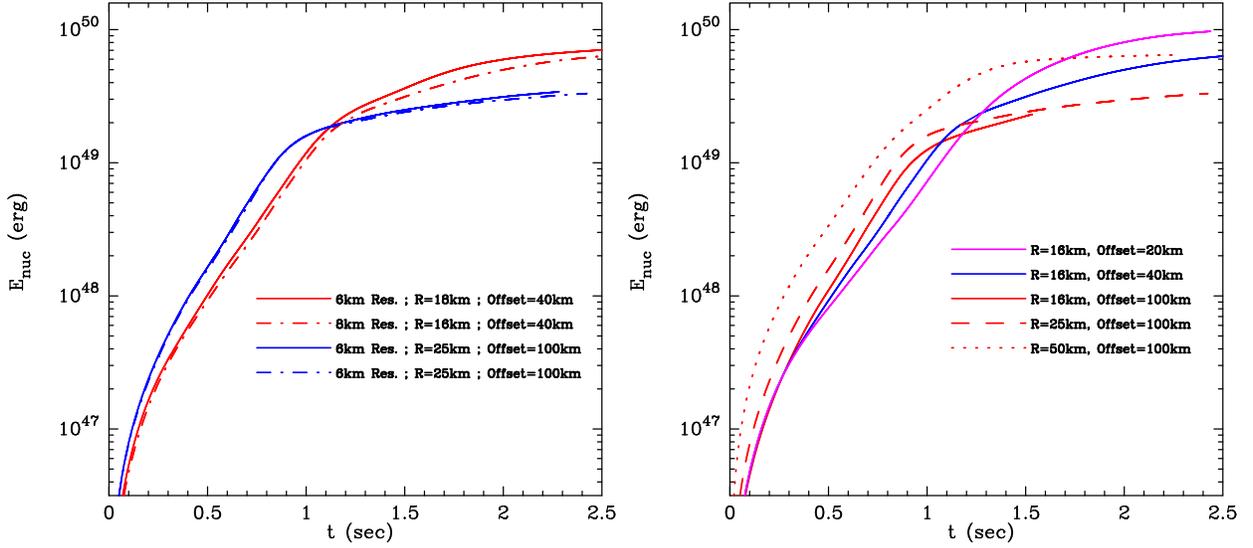

\includegraphics[angle=0,scale=0.43]{f4a.ps}
\includegraphics[angle=0,scale=0.43]{f4b.ps}
\caption{Nuclear energy released ($\enuc$) as a function of time. Left
panel:  Comparison of 6-km and 8-km resolution simulations for two
offset distances.  Note the close agreement between  the two
resolutions for both cases.  Right panel: Comparison of 8-km resolution
simulations for an initial bubble radius of 16-km and three different
initial offset distances (20, 40, and 100 km) and for two other initial
bubble radii (25 and 100 km) and an initial offset distance of 100 km. 
Note that the curves for initial bubble radii of 25 km, and especially
50 km -- initial conditions that violate the requirement for
self-similarity (see text) -- are displaced from the other curves, even
at early times.
\label{fig:energy_vs_time}}
\end{figure}

We compare the nuclear energy, $\enuc$, released during the deflagration
phase as a function of time for the 18b42o6r and 16b40o8r simulations,
and the 25b100o6r and 25b100o8r simulations, in the left panel of Figure
\ref{fig:energy_vs_time}.  In both cases, the curves for the two
different resolutions are in close agreement.  This result is expected
if buoyancy-driven nuclear burning depends mostly on the fluid dynamical
behavior at larger scales.  As remarked earlier, such a picture is
suggested by the results of \cite{zhang2007}, who find that the
time-averaged rate of buoyancy-driven nuclear burning did not vary when
the resolution was varied by a factor of four.  However, this is an open
question, as discussed in \S 2, and further studies are needed in order
to answer it.

We compare the nuclear energy released, $\enuc$, as a function of time
for the 16b20o8r, 16b40o8r, 16b100o8r, 25b100o8r, and 50b100o8r
simulations in the right panel of Figure \ref{fig:energy_vs_time}.  The
first three form a sequence in which both the time at which the curves
flatten and the amount of nuclear energy $\enuc$ that is released
decreases as the offset distance $\ro$ of the ignition region
increases.  This behavior is similar to that found by
\citet{townsley2007} in their 2-D simulations of the GCD model.

The unburnt surface material in the initial collision region reaches $T
> 3 \times 10^9$ K but densities of only $\rho \sim 10^5-10^6$ g
cm$^{-3}$, which are insufficient to produce a detonation \citep
{niemeyer1997}.  The collision, however, produces inward- and
outward-directed jets.  The outward jet ejects material at velocities
$\vjet \sim 40,000$ km s$^{-1}$.  The inward jet impacts the stellar
surface, stalls, and spreads a little.  This sequence of events
compresses the hot ($T > 3 \times 10^9$ K) material ahead of the jet to
densities $\rho > 1 \times 10^7$ g cm$^{-3}$, and in some cases even
$\rho > 2 \times 10^7$ g cm$^{-3}$.  These conditions exceed
conservative conditions for initiation of a detonation
\citep{niemeyer1997,roepke2007}.  As we will report in a later paper,
subsequent 3-D simulations of the GCD model we have carried out that
include nuclear burning outside the flame detonate, and inclusion of
such burning does not significantly alter the time at which the
detonation occurs compared to the first moment at which the simulations
reported in the present paper reach conditions for detonation
\citep{meakin2008b}.

\begin{figure}
\begin{center}
\includegraphics[scale=0.5,clip=]{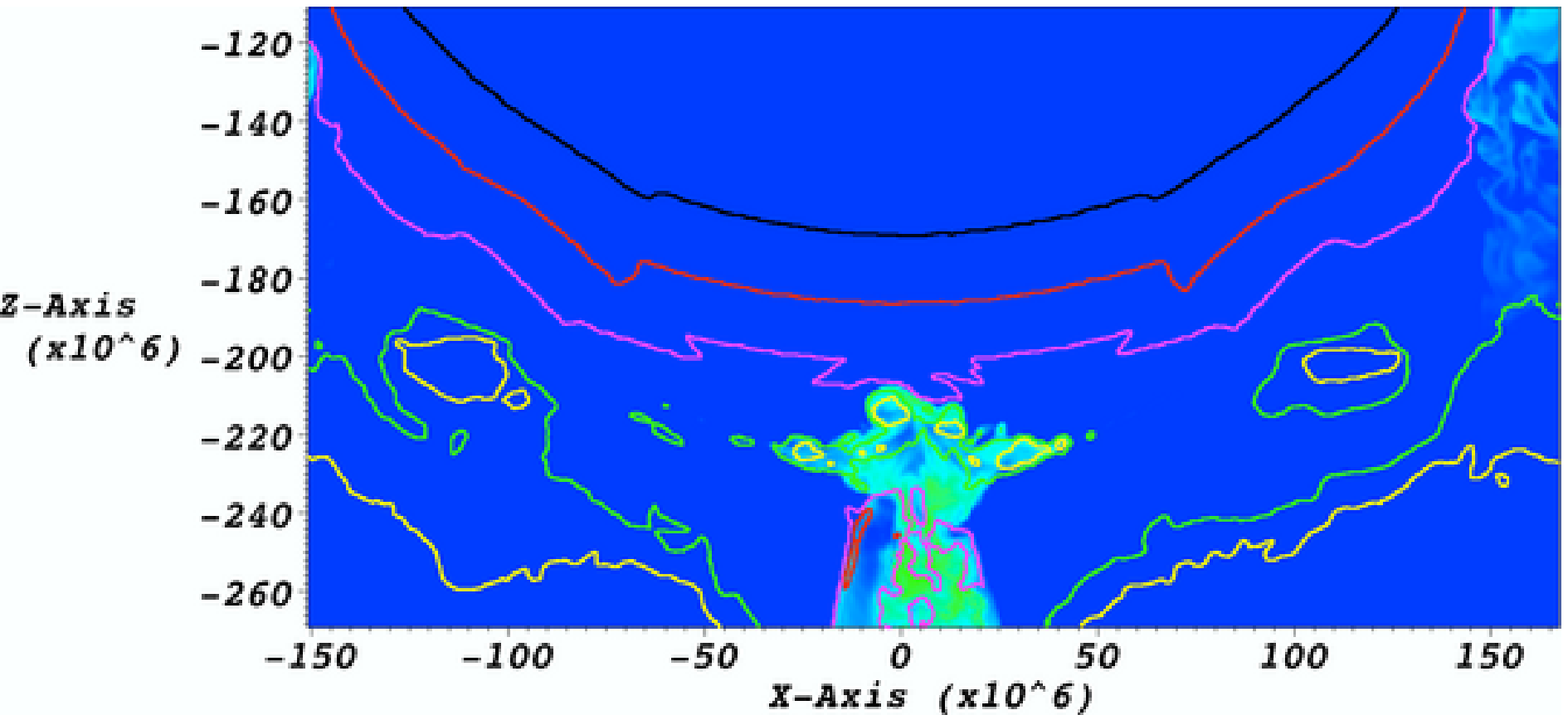}
\includegraphics[scale=0.5,clip=]{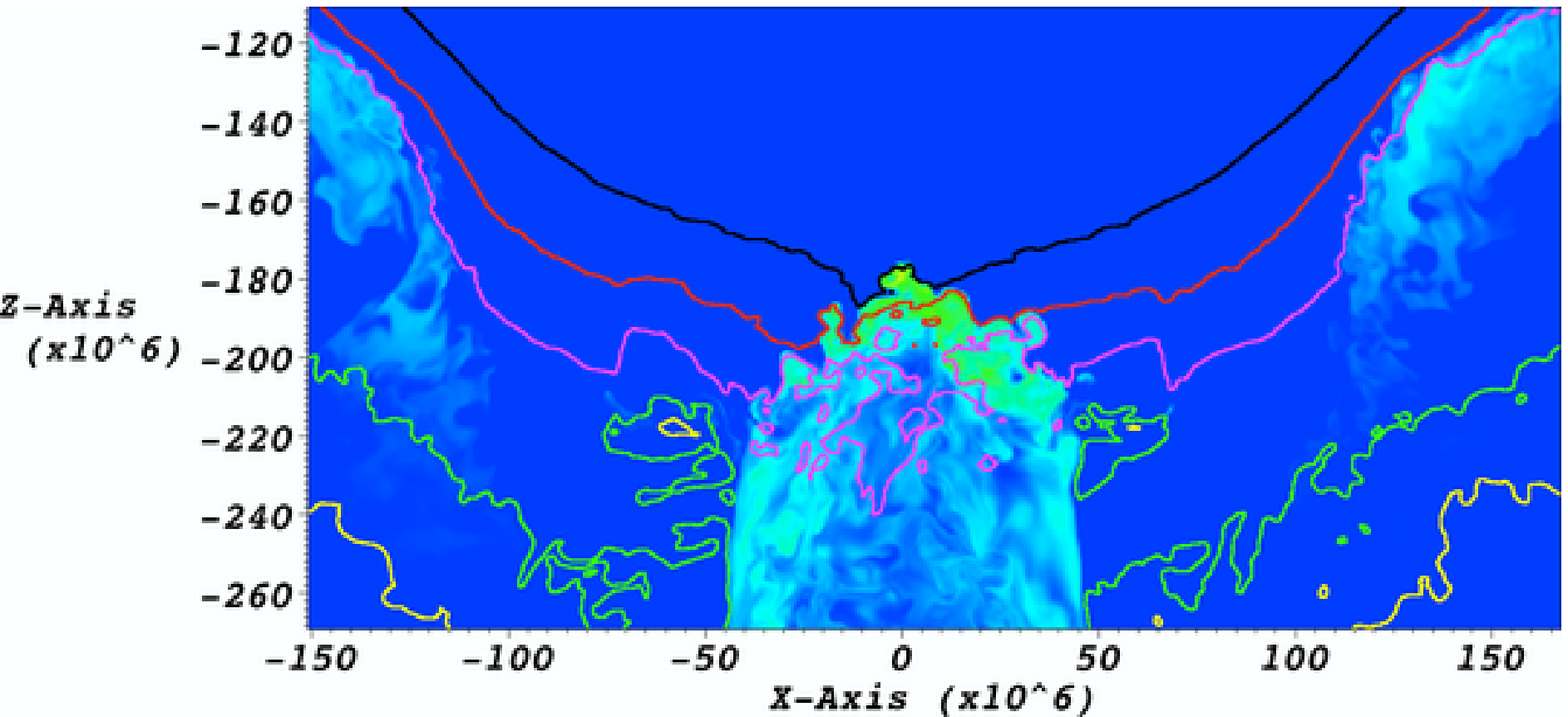}
\end{center}
\caption{Close-up view of 2-D slices of the region near the ``south
pole'' of the star.  The slices show the inward-directed jet produced
by the collision of unburnt material ahead of the hot ash from the
bubble in the 25b100o6r simulation just prior to when the density of
the material in the hot, inward-directed jet produced by the collision
has reached its maximum value.  The color shows the temperature,
ranging from  1 $\times 10^9$ - 5 $\times 10^9$ K from blue through
red. The density is indicated by contours.  The yellow contour
represents a density of $5 \times 10^5$ gm/cm$^3$, green $1 \times
10^6$ gm/cm$^3$, purple $5 \times 10^6$ gm/cm$^3$, red $1 \times 10^7$
g cm$^{-3}$, and black $2 \times 10^7$ gm/cm$^3$.
\label{fig:collision_region}}
\end{figure}

The 3-D simulations of the GCD model show that, as in our previous 2-D
simulations \citep{townsley2007}, it is the kinetic energy originating
from the breakout of the bubble of hot ash, imparted to the unburnt
surface layers of the star by the inwardly-moving jet generated by
collision of the surface flows that causes the unburnt material to
achieve the conditions for detonation.  This is illustrated in Figure
\ref{fig:collision_region}, which shows the temperature and density in
the collision region in the 18b42o6r simulation just before and just
after detonation conditions are reached, and in Figure
\ref{fig:hot_regions_rotated}, which is identical to the image in
Figure \ref{fig:hot_regions}(d), except that the star is rotated to
show the very hot ($T > 3 \times 10^9$ K) region at the head of the
inward jet that has reached densities greater than $2 \times 10^7$ g
cm$^{-3}$.

\begin{figure}
\includegraphics[angle=0,scale=0.86]{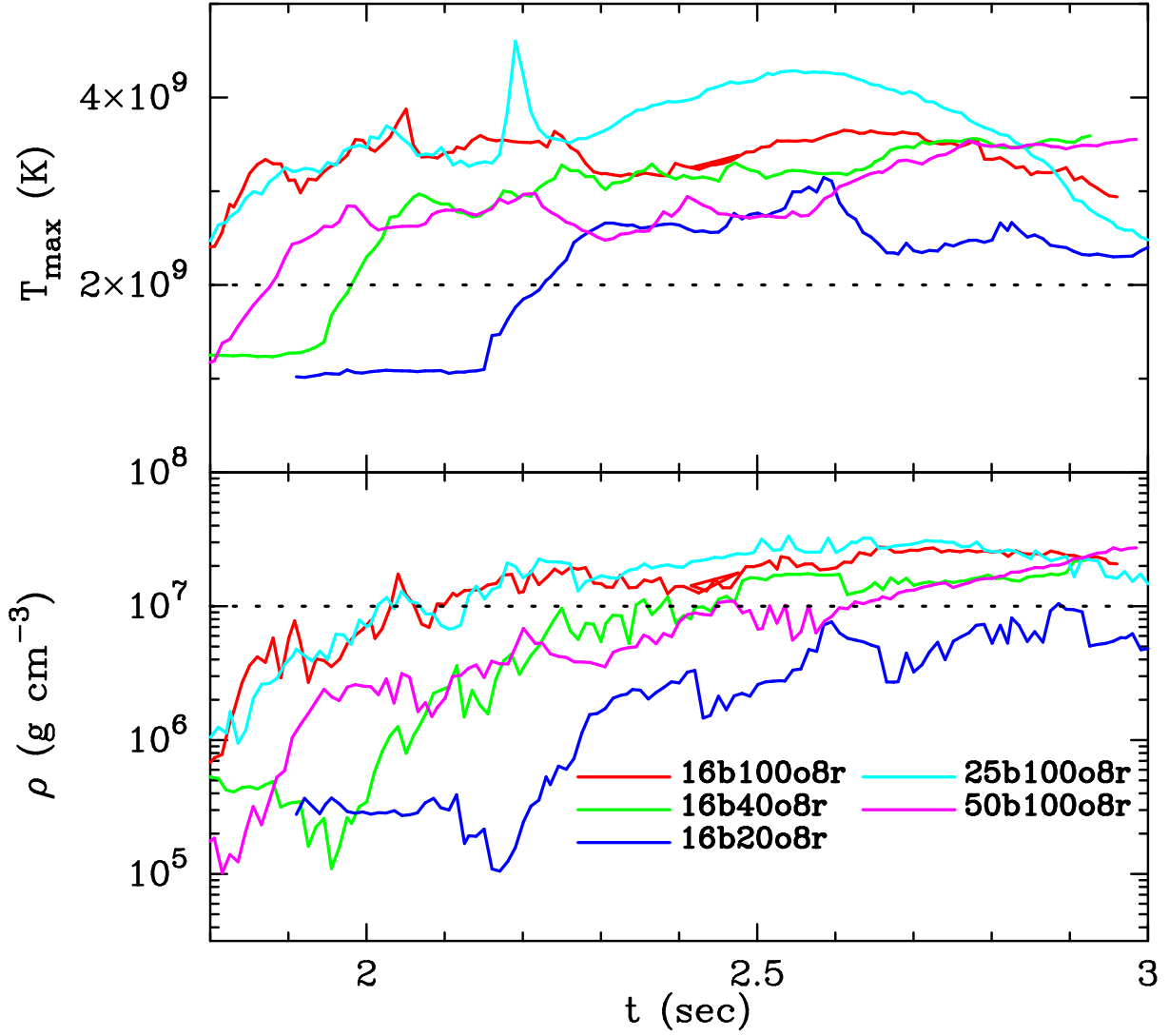}
\caption{Maximum temperature $T_{\rm max}$ and associated density
in the fully refined truncated cone around the ``south pole" of
the  star as a function of time for the five 8km resolution 3-D simulations we
performed.  The material flowing over the surface of the star enters the
lower hemisphere at $\sim$ 1.5 s and collides at $\sim$ 2 s, at which
point an inward-directed jet forms.  Subsequently, the hot ($T > 3
\times 10^9$ K) material in the jet impacts the surface of the star
and becomes compressed, reaching densities $\rho > 1 \times 10^7$ g
cm$^{-3}$ in all five of the simulations.
\label{fig:t_max_and_rho_max_vs_time}}
\end{figure}

We find that the unburnt material in the surface layers of the star
reaches temperatures $T > 2 \times 10^9$ K and densities $\rho > 1
\times 10^7$ g cm$^{-3}$ in all seven 3-D simulations we performed, as
illustrated in Figure~\ref{fig:t_max_and_rho_max_vs_time}.  The values
of $T_{\rm max}$ and $\rho_{\rm max}$ closely match those in our 2-D
cylindrical simulations for the same resolution and initial conditions
\citep{townsley2007}.  The small difference between the values of
$T_{\rm max}$ and $\rho_{\rm max}$ in the 2-D and 3-D simulations for
$\ro = 20$ km are within the uncertainties we expect in the
calculations.  Thus the results of our 2-D cylindrical simulations are a
good guide to the results of our 3-D simulations for the range of
initial bubble radii and offset distances, and the resolutions, that we
have explored so far.

In order to test the robustness of the GCD mechanism, we ran additional
simulations in which we coarsened the resolution in the truncated cone
encompassing the collision region from 8 km to 16 km, 32 km, and 64 km
for offset distances of 40 km and 100 km.  In all cases, the simulations
reached the above conservative conditions for detonation.  We conclude
that, for the initial conditions investigated in this paper, the GCD
model robustly achieves temperatures and densities necessary for
detonation.

\section{Discussion} \label{discussion}

\subsection{Observational Properties of the 3-D GCD Model}

We have carried out 3-D simulations of the GCD model at a finest
resolution of 6 km for initial offset distances of 42 km and 100 km, and
a finest resolution of 8 km for initial offset distances of 20 km, 42
km, and 100 km.  We find that these simulations robustly reach the
conditions necessary for detonation.  \citet{vladimirova2007} and
\citet{fisher2008} have shown that the evolution of the bubble is
self-similar -- i.e., its evolution is independent of its initial radius
$\rb$ -- provided that $\rb < \lc$, where $\rb$ is the initial bubble
radius and $\lc$ is the minimum wavelength for the unstable
Rayleigh-Taylor growth of flame surface perturbations (see \S 2).  We
find that the bubbles in the simulations in which $\rb = 25$ and 50 km
at an offset distance $\ro = 100$ km do not exhibit self-similar
behavior, even at early times; i.e., their size and shape at later times
differ from each other and from those in the simulation for which $\rb =
16$ km (see Figure \ref{fig:energy_vs_time}).  This is expected, since
in these two simulations the radius of the initial bubble does not
satisfy the condition $\rb < \lc$ required for self-similar behavior. 
These two simulations produced larger values of $\enuc$, but still
reached conservative conditions for detonation.  Finally, we find that
our 3-D simulations exhibit a correlation between $\enuc$ and initial
offset distance (see Figure \ref{fig:energy_vs_time}), confirming the
correlation seen in our 2-D cylindrical simulations
\citep{townsley2007}.  

Table 1 lists the amount of nuclear energy $\enuc$ released up to a
fiducial time of 2 s (which is approximately the time at which a
detonation would occur) for the seven simulations reported in this
paper.  Also listed in Table 1 are estimates of the masses of heavy
elements (i.e., iron-peak elements) and intermediate mass elements that
are expected to be produced by a subsequent detonation, assuming that
the mass of iron-peak elements $M_{\rm heavy}$ is that at densities
above $1.5 \times 10^7$ g cm$^{-3}$ and the mass of intermediate mass
elements $M_{\rm inter}$ is the mass at densities below this.  Finally,
Table 1 lists estimates of the expected total energy of the explosion,
$E_{\rm total} = \enuc + E_{\rm bind}$, where we estimate $\enuc$ from
the masses of heavy and intermediate mass elements that are produced in
each simulation, and $E_{\rm bind}$ is the binding energy of the
initial WD model.  These results suggest that all of the simulations
can be expected to produce large amounts of $^{56}$Ni, and therefore
very bright and energetic Type Ia supernova explosions, and small
amounts of intermediate mass elements.  Thus these simulations can
explain only the brightest and most energetic of the Type Ia supernovae
that are observed.

Simulation 50b100o8r, in which the initial bubble has a radius $\rb =
50$ km and therefore become immediately subject to a strong
Rayleigh-Taylor instability, may crudely mock up what happens if
ignition occurs simultaneously at a cluster of points located
off-center in the core of the star.  The simulation released more
$\enuc$ than did the other two simulations with the same offset
distance but with smaller initial bubble radii, and thus suggests a
plausible way in which the GCD mechanism could produce more
pre-expansion, and therefore much less nickel, yet detonate --
i.e., one way in which the GCD mechanism might account for less
luminous Type Ia supernovae.  However, this is an open question, which
we plan to explore in a future paper.

An essential aspect of the GCD model is that, while the nuclear energy
released during the deflagration phase causes the star to expand prior
to the detonation, it leaves the majority of the star unburnt and
undisturbed, as we have shown in earlier work
\citep{plewa2004,plewa2007,townsley2007} and in this work.  The
subsequent detonation phase can therefore be expected to produce a
smooth, stratified compositional structure in the interior of the star
similar to that inferred from spectroscopic observations of Type Ia
supernovae, and something that 2-D cylindrical and 3-D simulations of
both the pure deflagration model
[\citet{hoeflich2002,leonard2005,badenes2006,wang2006,wang2007}],
and the deflagration to detonation (DDT) model [see, e.g.
\citet{gerardy2007}] have difficulty in doing.  As is evident in
Figures \ref{fig:flame_surface} and \ref{fig:hot_regions}, the GCD model
also produces turbulence and compositional inhomogeneities in the
outermost layers of the star, which appear capable of matching
properties inferred from observations of line polarization in the
optical \citep{wang2006,wang2007} and line profiles in the NIR and MIR
\citep{gerardy2007}\footnote{It should be noted that one of the two
events discussed in \citep{gerardy2007} is a subluminous Type Ia
supernova, whereas the GCD simulations reported in this paper
produce very bright Type Ia supernovae.}.  Thus, the pure deflagration
and DDT models predict an inhomogeneous, mixed composition in the core
and a uniform composition in the outermost layers of the star -- a
compositional structure that is opposite to that inferred from
observations, while the GCD mechanism predicts a smoothly-stratified
composition in the core and an inhomogeneous, mixed composition in the
outermost layers of the star, which agrees qualitatively with the
compositional structure inferred from observations.

\subsection{Comparison With Other Work}

\citet{roepke2007} have recently conducted an extensive set of 2-D
cylindrical simulations and a few 3-D simulations of the deflagration
phase of the GCD model.  They find that the conditions for detonation
are reached for a number of their 2-D cylindrical simulations for a
variety of initial conditions.  However, the 3-D simulations they
performed did not reach conditions for detonation, whereas our 3-D
simulations do.

In an effort to understand the origin of this difference, we carried out
6-km and 8-km resolution simulations for exactly the same initial
conditions as were used for one of the two \citet{roepke2007} 3-D
simulations in which ignition was posited to occur at a single point: an
initial spherical bubble of radius 25 km offset a distance of 100 km
from the center of the star (see above).\footnote{We did not simulate
the other initial conditions for which \citet{roepke2007} did a 3-D
simulation positing a single ignition point (i.e., an initial bubble
radius of 25 km and an offset distance of 200 km) because these initial
conditions lie far above the $\rb = \lc$ curve, and are therefore not
physically self-consistent \citep{fisher2008}.} We also carried out an
8-km resolution simulation with an initial bubble radius $\rb = 50$ km
and an offset distance $\ro = 100$ km.  In all cases, the simulations
reached conservative conditions for detonation, as we have described
above.  The results provide evidence of the ability of the GCD model to
produce the conditions for detonation for a range of initial conditions,
but leave unanswered the question of why \citet{roepke2007} find that
the criteria for detonation are reached for a range of initial
conditions in their 2-D simulations but not in the 3-D simulations that
they performed, whereas we find that the criteria for detonation are
satisfied for a range of initial conditions in our 2-d simulations
\citep{townsley2007} and in our 3-D simulations, as reported in this
paper.

\begin{figure}
\includegraphics[angle=0,scale=1.2]{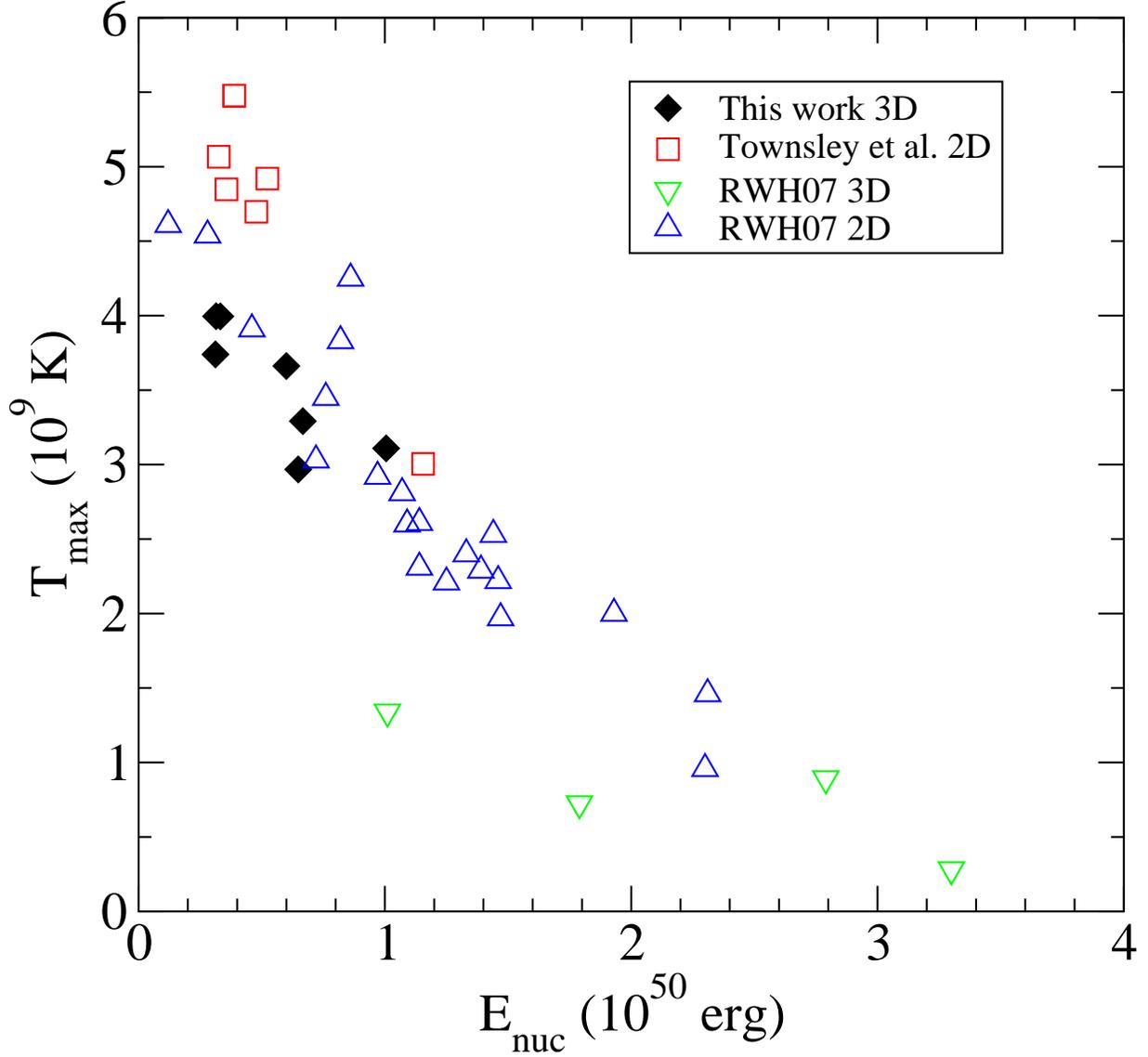}
\caption{Locations in the ($\enuc,\tmax$)-plane of our 2-D and 3-D
simulations of the GCD model and of R\"opke et al.'s 2-D and 3-D
simulations of the same model.  The four filled diamonds are (from left
to right) the locations of our 25b100o6r and 25b100o8r simulations
(for which the diamonds almost completely overlap) and our 16b40o8r and
18b42o6r simulations.  Note the correlation between $\enuc$ and $\tmax$
reported by \citep{roepke2007}.
\label{fig:t_max_vs_enuc_correlation}}
\end{figure}

We show the values of $\tmax$ and $\enuc$ for both the current 3-D
models and our previous 2-D models \citep {townsley2007}, and the 2-D
and 3-D models of \citet{roepke2007} in Figure
\ref{fig:t_max_vs_enuc_correlation}. This figure shows that there is a
relation between $\tmax$ in the collision region and $\enuc$.  Such a
relation is expected in the GCD model to the degree that larger values
of $\enuc$  produce more pre-expansion of the WD, and therefore less
kinetic energy in the flow of hot bubble material over the stellar
surface, leading to lower values of the temperature in the collision
region.  The amount of pre-expansion of the star can be expected to
depend on when the nuclear energy is released, as well as how much is
released.  The fact that the relation between $\tmax$ in the collision
region and $\enuc$ shown in Figure \ref{fig:t_max_vs_enuc_correlation}
is relatively narrow suggests that the amount of nuclear energy that is
released is the dominant factor in determining $\tmax$.

The results of our 2-D cylindrical and 3-D simulations for initial
conditions consisting of a single, small, spherical bubble offset a
range of distances from the center of the star agree with each other,
as previously noted, and lie on the relation between $\tmax$ and
$\enuc$.  So do the results of R\"opke et al.'s 2-D cylindrical
simulations for initial conditions consisting of a cluster of bubbles
and for two tear-drop-shaped ignition regions located on opposite sides
of the center of the star, as well as a single spherical bubble,
all offset a range of distances from the center of the star.  The
results of their 3-D simulations for initial conditions consisting of a
single bubble also follow the relation between $\tmax$ and $\enuc$,
but those for initial conditions consisting of a cluster of bubbles and
for two tear-drop-shaped ignition regions located on opposite sides of
the center of the star do not.  In particular, their 3-D simulations
starting with a cluster of bubbles release a low enough $\enuc$ that
they should reach detonation conditions if the relation between $\tmax$
and $\enuc$ were followed.  Most importantly, \citet{roepke2007} make
no mention of the outward-directed jet, and especially the
inward-directed jet, which we find plays a crucial role in achieving
detonation conditions.  Consequently, it is difficult to make direct
comparisons between our simulations and theirs.

The main conclusion we draw from Figure
\ref{fig:t_max_vs_enuc_correlation} is that the amount of $\enuc$
released in R\"opke et al.'s 2-D simulations is more, and in R\"opke et
al.'s 3-D simulations is much more, than is released in our 2-D
\cite{townsley2007} and 3-D simulations, which release similar amounts
of $\enuc$ for the same initial conditions and resolution. 
Consequently, the kinetic energy of the surface flow in their 3-D
simulations -- as measured by $\tmax$ -- is much smaller and the
simulations do not achieve detonation conditions. 

Without knowledge of the details of their simulations, it is difficult
to know why the results of their 3-D simulations differ from those of
their 2-D simulations, which -- while releasing more $\enuc$ -- lead to
a surface flow similar to what we see in our 2-D and 3-D simulations. 
A likely reason is the different treatments of buoyancy-driven
turbulent nuclear burning in their simulations and in ours.  Our
treatment assumes that the rate of buoyancy-driven turbulent nuclear
burning depends mostly on the behavior of the flow at larger scales. As
we have seen, this provides an explanation for why the results of our
2-D and 3-D simulations agree.  It may also provide an explanation for
why the results of R\"opke et al.'s 2-D simulations agree with our 2-D
(and therefore also our 3-D) simulations:  their treatment of
buoyancy-driven turbulent nuclear burning uses the properties of the
flow at scales above the grid scale of the simulation to determine
the turbulent energy at subgrid scales, and therefore the increase in
the nuclear burning rate (which is parameterized as an increase in the
value of the flame speed) due to this turbulence.  The fact that
turbulence in 3-D leads to a cascade of smaller and smaller eddies,
while turbulence in 2-D does not, means that their treatment of
buoyancy-driven turbulent nuclear burning does not increase the rate
of nuclear burning in 2-D, whereas in 3-D it will.  The origin of the
similarities between our 2-D and 3-D results and R\"opke et al.'s 2-D
results, and the difference between these results and R\"opke et al.'s
3-D results, are thus most likely due to differences in the treatment
of buoyancy-driven turbulent nuclear burning.  As we have noted above,
the appropriate treatment of such burning is an open question, and is --
as the above differences emphasize -- an important topic for future
study.

\section{Conclusions} \label{conclusions}

We have conducted a series of 3-D simulations of the GCD mechanism for
several offset distances and resolutions.  Conservative conditions
necessary for detonation are robustly achieved in all cases.  The
initial conditions explored in this paper lead to conditions at
detonation that can be expected to produce large amounts of $^{56}$Ni
and small amounts of intermediate mass elements.  These particular
simulations are therefore relevant only to high luminosity Type Ia
supernovae.  We find a correlation between the central density of the
star at detonation and both the offset distance and the radius of the
initial bubble.  These correlations offer a possible explanation for the
observed variation in nickel mass in Type Ia supernovae.  Finally, the
uniform, homogeneous cores and the turbulent, heterogeneous composition
of the outer layers of the stars at the time when the conditions for
detonation are reached match the properties inferred from recent
polarization, NIR, and MIR observations of Type Ia supernovae.

\acknowledgments
The authors thank Nathan Hearn for discussions of this work and for his
software analysis tools, which were invaluable in the preparation of
this paper.  The authors thank the code group in the Flash Center,
especially Anshu Dubey, Lynn Reid, Paul Rich, Dan Sheeler, and Klaus
Weide, for development of the code and for help in running our large
simulations on uP at LLNL.  We also thank Brad Gallagher and the
visualization group in the Flash Center for creating the images used in
Figures 1, 2, and 3, and the corresponding movies; and Hank Childs and
the VisIt team at LLNL for help in using VisIt.  Finally, we thank LLNL
Computing for help in running our large simulations on uP at LLNL, and
the NERSC support staff at LBNL for help in running them on Bassi and
Seaborg.  This work is supported in part at the University of Chicago
by the U.S Department of Energy (DOE) under Contract B523820 to the ASC
Alliances Center for Astrophysical Nuclear Flashes, and in part by the
National Science Foundation under Grant PHY 02-16783 for the Frontier
Center ``Joint Institute for Nuclear Astrophysics'' (JINA).  This
research used computational resources awarded under the INCITE program
at LBNL NERSC, which is supported by the Office of Science of the U.S.
Department of Energy under Contract No. DE-AC03-76SF00098.  ACC
acknowledges support from NSF Grant ST-0507456.  JWT acknowledges
support from Argonne National Laboratory, which is operated under DOE
Contract No. W-31-109-ENG-38. 


\clearpage

\begin{deluxetable}{ccccccccc}
\tablecaption{Properties of 3-D GCD Simulations\label{tbl:location}}
\tablewidth{0pt}
\tablehead{
\colhead{Label} &
\colhead{$\rb$} &
\colhead{$\ro$} &
\colhead{Resolution} & 
\colhead{$E_{\rm nuc,def}$} &
\colhead{$M_{\rm heavy}$} &
\colhead{$M_{\rm inter}$} &
\colhead{$E_{\rm total}$} &
\colhead{$t_{\rm det}$} \\
\colhead{} & 
\colhead{(km)} & 
\colhead{(km)} &
\colhead{(km)} & 
\colhead{($10^{49}{\rm \ erg}$)} & 
\colhead{($M_\sun$)} & 
\colhead{($M_\sun$)} & 
\colhead{($10^{51}{\rm \ erg}$)} &
\colhead{(s)}
}
\startdata
16b20o8r    & 16  & 20   & 8  & 10.5 & 1.00  & 0.36  & 1.50  & 2.89  \\
18b42o6r    & 18  & 42   & 6  & 6.7  & 1.18  & 0.19  & 1.57  & 2.30  \\
16b40o8r    & 16  & 40   & 8  & 6.1  & 1.20  & 0.16  & 1.58  & 2.38  \\
16b100o8r   & 16  & 100  & 8  & 3.2  & 1.26  & 0.10  & 1.60  & 2.02  \\
25b100o6r   & 25  & 100  & 6  & 3.0  & 1.27  & 0.17  & 1.60  & 1.84  \\
25b100o8r   & 25  & 100  & 8  & 3.1  & 1.26  & 0.11  & 1.60  & 2.01  \\
50b100o8r   & 50  & 100  & 8  & 6.5  & 1.11  & 0.25  & 1.55  & 2.45  \\
\enddata
\vskip -18pt
\tablecomments{This table gives the properties of the seven 3-D
simulations of GCD models reported in this paper.}
\end{deluxetable}

\clearpage

\begin{thebibliography}{999}

\bibitem[Asida et al.(2008)]{asida2008}
	Asida, S., et al. 2008, ApJ, to be submitted

\bibitem[Badenes et al.(2006)]{badenes2006}
	Badenes, D. C., et al. 2006, \apj, 645, 1373

\bibitem[Bravo et al.(2006)]{bravo2006}
	Bravo, E., \& Garcia-Senz, D. 2006, ApJ, 642, L157

\bibitem[Calder et al.(2002)]{calder2002} 
	Calder, A.~C., et al. 2002, \apjs, 143, 201 

\bibitem[Calder et al.(2004)]{calder2004}
	Calder, A. C., et al. 2004, astro-ph/0405162

\bibitem[Calder et al.(2007)]{calder2007}
 	Calder, A. C., et al. 2007, \apj, 656, 313

\bibitem[Fisher et al.(2008)]{fisher2008}
	Fisher, R., et al. 2008, \apj, to be submitted

\bibitem[Fryxell et al.(2000)]{fryxell2000}
	Fryxell, B., et al. 2000, APJS, 131, 273

\bibitem[Gamezo et al.(2003)]{gamezo2003}
 	Gamezo, V. N., et al. 2003, Science, 299, 77

\bibitem[Gamezo, Khokhlov \& Oran(2004)]{gamezo2004}
	Gamezo, V. N., Khokhlov, A. M., \& Oran, E. S.  2004, \prl, 92,
	211102

\bibitem[Gamezo et al.(2005)]{gamezo2005}
	Gamezo, V. N., Khokhlov, A. M., \& Oran, E. S.  2005, \apj, 623,
	337

\bibitem[Gerardy et al.(2007)]{gerardy2007}
	Gerardy, C., et al. 2007, \apj, 661, 995

\bibitem[H\"oflich et al.(2002)]{hoeflich2002}
	H\"oflich, P., et al. 2002, \apj, \apj, 568, 791


\bibitem[Jordan et al.(2008)]{meakin2008b}
	Jordan, G. C., Meakin, C., A., Seitenzahl, I., Townsley, D. M.,
	Fisher, R. T., Graziani, C., Truran, J. W., \& Lamb, D. Q.
	2008b. in preparation

\bibitem[Khokhlov(1991)]{khokhlov1991}
	Khokhlov, A. M. 1991, A\&A, 245, 114 

\bibitem[Khokhlov(1995)]{khokhlov1995}
	Khokhlov, A. M. 1995, \apj, 449, 695


\bibitem[Leonard et al.(2005)]{leonard2005}
	Leonard, D C., et al. 2005, \apj, 632, 450

\bibitem[Livne et al.(2005)]{livne2005}
	Livne, E., Asida, S . M., \& H\"oflich, P.  2005, \apj, 632, 443

\bibitem[Meakin et al.(2008a)]{meakin2008a}
	Meakin, C., A., Seitenzahl, I., Townsley, D. M., Jordan, G.,
	Truran, J., \& Lamb, D. Q. 2008a. \apj, to be submitted

\bibitem[Niemeyer(1999)]{niemeyer1999}
	Niemeyer, J. C. 1999, \apj, 523, L57

\bibitem[Niemeyer \& Woosley(1997)]{niemeyer1997}
	Niemeyer, J. C. \& Woosley, S. E. 1997, \apj, 475, 740

\bibitem[Nomoto, Thielemann, \& Yokoi(1984)]{nomoto1984}
	Nomoto, K., Thielemann, F.-K., \& Yokoi, K. 1984, \apj, 286, 644

\bibitem[Perlmutter et al.(1998)]{perlmutter1998}
        Perlmutter, S., et al. 1998, Abstracts of the 19th Texas
        Symposium on Relativistic Astrophysics and Cosmology, held in
        Paris, France, Dec. 14-18, 1998. Eds.: J. Paul, T. Montmerle,
        and E. Aubourg (CEA Saclay)

\bibitem[Plewa(2007)]{plewa2007}
	Plewa, T. 2007, \apj, 657, 942

\bibitem[Plewa, Calder, \& Lamb(2004)]{plewa2004}
	Plewa, T., Calder, A. C., \& Lamb, D. Q. 2004, \apj, 612, L37

\bibitem[Reinecke et al.(1999)]{reinecke1999}
	Reinecke, M., et al. 1999, A\&A, 347, 724

\bibitem[Reinecke et al.(2002a)]{reinecke2002a}
 	Reinecke, M., et al. 2002a, A\&A, 386, 936

\bibitem[Reinecke et al.(2002b)]{reinecke2002b}
 	Reinecke, M., et al. 2002b, A\&A, 391, 1167

\bibitem[Riess et al.(1998)]{riess1998}
	Riess, A., et al. 1998, AJ, 116, 1009

\bibitem[R\"opke, Niemeyer, \& Hillebrandt(2006a)]{roepke2003}
	R\"opke, F., Niemeyer, J.C., \& Hillebrandt, W. 2003, \apj, 588,
	952

\bibitem[R\"opke \& Hillebrandt(2005)]{roepke2005}
 	R\"opke, F., \& Hillebrandt, W. 2005, A\&A, 431, 635

\bibitem[R\"opke, Woosley, \& Hillebrandt(2007)]{roepke2007}
	R\"opke, F., Woosley, S. E., \& Hillebrandt, W. 2007, 660, 1344

\bibitem[Schmidt, Niemeyer, \& Hillebrandt(2006a)]{schmidt2006a}
	Schmidt, W., Niemeyer, J.C., \& Hillebrandt, W. 2006, A\&A 450,
	265

\bibitem[Schmidt, Niemeyer, \& Hillebrandt(2006b)]{schmidt2006b}
	Schmidt, W., Niemeyer, J.C., \& Hillebrandt, W. 2006, A\&A 450,
	283

\bibitem[Timmes \& Woosley(1992)]{timmes1992}
	Timmes, F. X. \& Woosley, S. E. 1992, \apj, 396, 649

\bibitem[Townsley et al.(2007)]{townsley2007}
	Townsley, D., et al. 2007, \apj, 668, 1118

\bibitem[Townsley et al.(2008)]{townsley2008}
	Townsley, D., et al. 2008, \apj, to be submitted

\bibitem[Vladimirova(2007)]{vladimirova2007}
	Vladimirova, N. 2007, Combust. Theory Modeling, 11, 377

\bibitem[Vladimirova, Weirs, \& Ryzhik(2006)]{vladimirova2006}
	Vladimirova, N., Weirs, G., \& Ryzhik, L. 2006, Combust. Theory
	Modeling, 10, 727

\bibitem[Wang et al.(2006)]{wang2006}
	Wang, L., et al. 2006, \apj, 653, 490

\bibitem[Wang et al.(2007)]{wang2007}
	Wang, L., et al. 2007, Science, 315, 212

\bibitem[Woosley, Wunsch, \& Kuhlen(2004)]{woosley2004}
	Woosley, S. E., Wunsch, S., \& Kuhlen, M. 2004, \apj, 607, 921

\bibitem[Wunsch \& Woosley(2004)]{wunsch2004}
	Wunsch, S. \& Woosley, S. E. 2004, \apj, 616, 1102

\bibitem[Zhang et al.(2007)]{zhang2007}
	Zhang, J., et al. 2007, \apj, 656, 347

\bibitem[Zingale et al.(2005)]{zingale2005}
	Zingale, M., et al. 2005, \apj, 632, 1021


\end{thebibliography}
\end{document}